# Multimodal Digital Sensing of Early-Life Laying Hens: A Pilot Study Integrating Thermal, Acoustic, Optical-Flow and Environmental Data

Yashan Dhaliwal[1], Daniel Essien[2], Suresh Neethirajan[1,2,*]

[1]Faculty of Agriculture, Dalhousie University, Truro, NS B2N 5E3, Canada
[2]Faculty of Computer Science, Dalhousie University, Halifax, NS B3H 1W5, Canada
*Corresponding author. Email: sneethir@gmail.com

**Abstract**
Early-life development profoundly shapes long-term welfare in laying hens, yet monitoring remains constrained by subjective assessment and fragmented single-modality tools. This pilot study evaluated the technical feasibility of a multimodal sensing approach integrating thermal imaging, acoustic recording, optical-flow-based video analysis, and environmental monitoring to characterize physiological and behavioural development from hatch to 20 weeks. One hundred fifty Lohmann LSL-Lite chicks were housed across five controlled rooms; thermal and environmental data were collected system-wide, whilst detailed audio and video analyses focused on one representative room to manage annotation workload. Weekly aggregated features included head and foot surface temperatures, acoustic spectral descriptors, optical-flow movement metrics around caretaker entry, and ambient conditions. Thermal imaging revealed age-related increases and stabilization of peripheral temperatures, with foot temperature showing a pronounced developmental effect ($\eta^2 = 0.51$). Acoustic features shifted systematically across weeks (all $p < 0.001$), consistent with vocal maturation. Optical-flow analysis revealed strong early reactivity to caretaker presence that declined markedly with development (early weeks 5–10 vs. late weeks 11–20: $t = 28.12$, $p = 0.00126$). Z-score-normalized multimodal trajectories and Pearson correlation analysis (Benjamini-Hochberg FDR, $q < 0.05$) demonstrated strong within-modality consistency ($r = 0.85–0.96$) and selective associations between environmental humidity and acoustic features ($r = 0.65–0.70$), whilst thermal, acoustic, and behavioural domains remained largely independent. This descriptive pilot establishes baseline multimodal developmental patterns and validates parallel sensing as a foundation for future welfare-relevant monitoring in precision poultry farming.

## 1. Introduction

*1.1 The critical juncture of early-life development in laying hens*
Early-life development in laying hens, spanning the brooding and rearing phases from hatch to approximately 20 weeks of age, constitutes a critical biological window that shapes long-term health, productivity, and behavioural stability into adulthood. Physiological and behavioural trajectories established during this period exert substantial and well-documented influence on subsequent laying performance, skeletal integrity, immune competence, and resilience to chronic disease across the productive lifespan (1-5). Early-life conditions affect not only survival and growth, but also the formation of fear responses, social dominance hierarchies, and adaptive capacity to environmental perturbations (6-8). These phenotypic characteristics remain remarkably stable into adulthood and directly influence flock-level productivity, welfare outcomes, and economic viability in commercial systems.

Despite this recognized importance, welfare-relevant characterization of early-life development in laying hens remains fragmented and incomplete. Monitoring practices in both research and commercial contexts continue to rely primarily on intermittent visual inspection and subjective scoring systems (9). Such approaches are constrained by low temporal resolution, observer variability, and an inherent inability to detect subtle physiological or behavioural shifts before they manifest as overt clinical or production-level outcomes (10). Visual assessments are typically conducted at irregular intervals, often weekly or less, thereby missing transient physiological events and gradual developmental inflection points that may be critical for welfare outcomes (9, 11). At the same time, the early-life rearing environment is highly dynamic, shaped by substantial thermal variation, rapidly evolving social structure as dominance hierarchies emerge, and repeated human–animal interactions that differ in frequency, consistency, and aversiveness depending on management intensity and caretaker behaviour. Yet systematic longitudinal descriptions of how individual birds and flocks adapt physiologically and behaviourally to these changing conditions remain scarce in the peer-reviewed literature, particularly at a mechanistic level.

*1.2 Limitations of single-modality monitoring in current practice*
Contemporary poultry welfare assessment remains largely unidimensional. Environmental conditions such as temperature, humidity, air quality, and light intensity may be monitored continuously and in detail, yet often in isolation from behavioural observation or physiological context. Conversely, behavioural assessments derived from ethological scoring or activity sensors are frequently disconnected from the environmental and physiological drivers shaping observed patterns. Even with the emergence of precision livestock farming technologies over the past decade, including networked environmental sensors, computer vision systems, and acoustic monitoring tools, most deployed and reported systems continue to operate independently, with minimal integration across modalities (12-15).

This fragmentation has predictable consequences for biological interpretation. Video-based computer vision approaches can quantify movement patterns, spatial distribution, and activity levels with high precision, yet offer limited insight into thermoregulatory state or internal physiological stress (16, 17). Acoustic monitoring systems can sensitively detect changes in vocal behaviour associated with arousal, social interaction, or distress, but cannot resolve concurrent locomotor responses or environmental tolerance thresholds (18-20). Thermal imaging and infrared thermography provide valuable proxies for heat balance and peripheral blood flow, yet cannot directly infer behavioural adaptation or fear responses (21, 22). Environmental sensors capture physical conditions accurately, but provide no information on how birds perceive, tolerate, or behaviourally respond to environmental perturbations (23, 24).

This modality-specific limitation constrains interpretive power and fundamentally restricts the capacity of single-modality systems to capture the multidimensional nature of early-life development (25, 26). Welfare does not arise from a single domain, but emerges from interactions among physiological homeostasis, behavioural expression, environmental appraisal, and social context (27). Commercial adoption of comprehensive multimodal monitoring frameworks has been further impeded by implementation costs, demanding data-processing requirements, and persistent uncertainty regarding how heterogeneous sensor streams with differing temporal resolution and information content should be meaningfully combined into biologically interpretable indicators (18, 27). Consequently, there remains a notable absence of peer-reviewed studies that systematically integrate multiple sensing modalities operating in parallel to track coordinated developmental change in laying hens across early life. This gap represents both a scientific opportunity and a practical barrier to objective, technology-enabled welfare monitoring.

*1.3 Rationale for multimodal integration*
Multimodal sensing provides a biologically grounded approach to addressing these methodological and interpretive limitations (28). By capturing complementary aspects of shared physiological and behavioural processes, multiple independent sensor systems can provide convergent evidence of developmental state and welfare trajectory (18, 27, 28). Thermoregulatory maturation, for example, should be reflected not only in stabilizing body surface temperatures, but also in modulated vocal characteristics and altered behavioural responsiveness to environmental or human-related stimuli. Environmental variability may elicit coordinated, though not identical, responses across physiological and behavioural domains, with differing timescales and amplitudes depending on underlying mechanisms. When multiple modalities demonstrate strong internal consistency alongside selective cross-modal coherence, confidence in biological interpretation is strengthened and the likelihood of measurement artefact is reduced. Conversely, selective dissociation between modalities may reveal independent developmental pathways or genuine welfare perturbations that single-modality systems would fail to detect.

Importantly, the value of multimodal approaches does not depend on sophisticated real-time artificial intelligence or complex data fusion at the pilot stage. Descriptive characterization of how thermal, acoustic, behavioural, and environmental trajectories co-vary over developmental time provides essential evidence of biological coherence and measurement fidelity. Such baseline mapping is a necessary precursor to hypothesis-driven experimentation, supervised learning, and eventual deployment of automated monitoring systems in commercial settings. For laying hens, while isolated systems targeting thermal comfort, acoustic distress, movement dynamics, or environmental parameters have been reported, no published study has integrated all four modalities in parallel to comprehensively characterize development from hatch through point-of-lay.

*1.4 Study design and objectives*
In this context, we conducted a pilot longitudinal study to establish baseline multimodal developmental trajectories and to assess technical feasibility in early-life laying hens housed in a controlled research facility. Four complementary sensing modalities were deployed in parallel: thermal imaging, acoustic recording, optical-flow-based video analysis, and environmental monitoring. Developmental patterns were tracked from hatch to 20 weeks of age, encompassing the transition from brooding through rearing toward sexual maturity.

Thermal imaging and environmental monitoring were conducted across all five experimental rooms to establish system-wide comparability. Detailed audio and video analyses were intentionally restricted to one representative room due to the substantial manual workload associated with frame-level video processing and clip-based acoustic annotation. This scope restriction was defined a priori during study design and reflects feasibility assessment rather than a limitation of experimental design. Such asymmetry is methodologically appropriate for a pilot study and consistent with early-stage multimodal sensing research.

The specific objectives were to:
1. Characterize age-related trajectories in surface thermoregulation, vocal spectral features, and flock movement responses to routine caretaker entry across the 20-week developmental window.
2. Evaluate the technical feasibility of parallel multimodal data acquisition, processing, and storage over extended developmental periods, identifying practical bottlenecks and scalability constraints relevant to future commercial deployment.
3. Quantify cross-modal associations between thermal, acoustic, behavioural, and environmental features to assess whether these modalities capture interrelated or independent dimensions of early-life development.

This study is explicitly descriptive in scope. It does not attempt to classify welfare states, establish causal inference, generate real-time alerts, or validate sensor outputs against independent physiological welfare biomarkers. Instead, it establishes empirically grounded baseline developmental patterns and demonstrates the technical feasibility of parallel multimodal sensing, providing essential scaffolding for future welfare validation, biomarker integration, and predictive modelling in precision poultry farming.

## 2. Materials and Methods
### 2.1 Study design and scope
This study employed a longitudinal observational design to characterise developmental patterns in early-life laying hen physiology and behaviour using multimodal sensor data. A total of 150 Lohmann LSL-Lite chicks were monitored from hatch to 20 weeks of age across five controlled environmental rooms, representing the full brooding and rearing period up to sexual maturity. Four complementary sensing modalities were deployed in parallel: thermal imaging, acoustic recording, optical-flow-based video analysis, and environmental monitoring.

A deliberate methodological asymmetry was incorporated at the study design stage. Thermal imaging and environmental monitoring were conducted across all five rooms to establish replicate-level comparability and to capture system-wide developmental trends. In contrast, detailed video and acoustic analyses, which require frame-level or clip-based manual annotation, were intentionally restricted to one representative room. This restriction was defined a priori and reflects a realistic assessment of manual processing demands inherent in early-stage multimodal monitoring. Such asymmetry is methodologically appropriate for a pilot feasibility study focused on demonstrating technical integrability and biological coherence rather than achieving statistical power across all modalities. The single room focus for audio and video therefore represents an explicit delineation of analytical depth rather than a limitation of experimental design.

### 2.2 Experimental housing and husbandry
One hundred fifty Lohmann LSL-Lite chicks, a commercial layer strain widely used in global egg production, were housed and reared in five identical controlled environmental rooms at the Atlantic Poultry Research Centre, Dalhousie Agricultural Campus, Truro, Nova Scotia. Each room housed 30 chicks under standardised rearing conditions, including graduated heating protocols with initial brooding temperatures of approximately 35°C, declining to approximately 21°C by week 4, mechanical ventilation, and a 16 h:8 h light: dark photoperiod aligned with commercial practice.

Birds were group-housed on wood-shaving litter flooring and provided ad libitum access to water and commercially formulated starter and grower diets appropriate to developmental stage, in accordance with institutional animal care guidelines. Routine husbandry practices, including feeding schedules, health inspections, vaccination procedures, and equipment maintenance, were carried out by trained caretaking staff. Daily husbandry records, including timing of caretaker entry and identification of non-routine events, were systematically documented and later used as contextual metadata when interpreting behavioural, acoustic, or thermal anomalies.

## 2.3 Multimodal sensor deployment and data acquisition
### 2.3.1 Thermal imaging

Radiometric thermal images were acquired every third day from each of the five rooms using a calibrated FLIR A-series thermal camera operated in radiometric mode. Images were captured manually at a standardized distance of 1 m and consistent camera angles to ensure temporal comparability and minimize confounding effects of acquisition geometry. Embedded radiometric data were extracted using the FLIR Ignite cloud-based platform. Figure 1 illustrates the radiometric thermal imaging workflow, showing the calibrated FLIR camera setup and a representative annotated thermal image used to extract head and foot surface temperatures.

For each image, minimum, maximum, and mean temperatures were recorded separately for head and foot regions of interest using manual annotation tools. Images affected by motion artefact, subject occlusion, or poor focus were excluded through visual quality control. This procedure yielded approximately 120 to 140 high-quality thermal images per room across the 20-week study period, corresponding to approximately 6 to 7 images per week, and enabled robust longitudinal characterisation of thermoregulatory development.

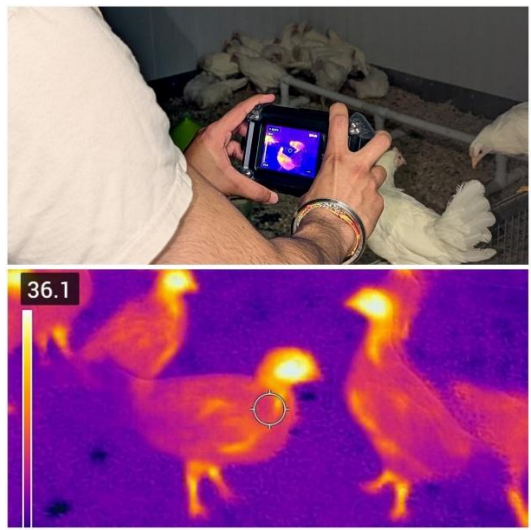

Figure 1. Radiometric thermal imaging workflow. (Upper) Calibrated FLIR A-series thermal camera used for manual image acquisition at a standardised distance of 1 m. (Lower) Representative thermal image

illustrating spatial temperature distribution, with head and foot regions of interest annotated for temperature extraction.

### 2.3.2 Acoustic recording

Audio data were collected using four complementary recording systems deployed across four of the five rooms. Detailed analysis focused on Room 1, while recordings from Rooms 2 to 4 were retained for future comparative and scaling analyses. The acoustic array comprised two Zoom H4n Pro recorders fitted with external RODE lavalier microphones, one Zoom F6 professional field recorder with an external RODE microphone, and one Wildlife Acoustics Song Meter SM4 autonomous recorder. Figure 2 depicts the acoustic recording systems used in the experimental rooms, including Zoom recorders with external RODE microphones and a Song Meter SM4 unit, all mounted at consistent heights to standardize whole-room vocalisation capture.

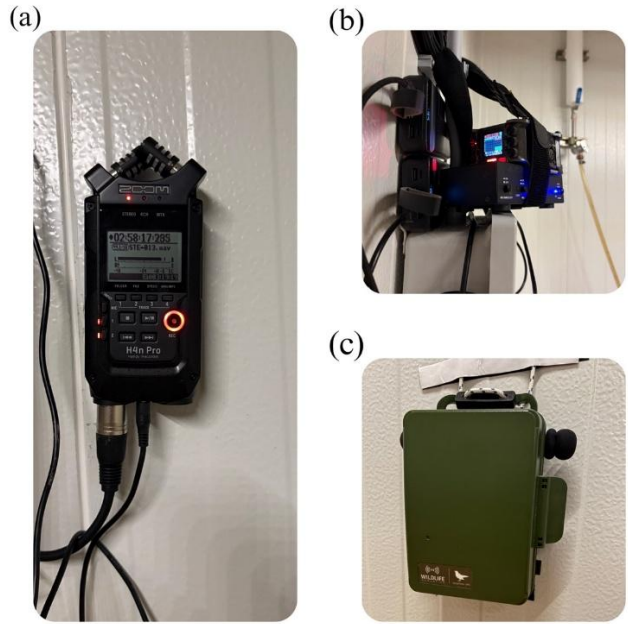

Figure 2. Acoustic recording systems deployed in experimental rooms. (a) Zoom H4n Pro recorder with external RODE lavalier microphone. (b) Zoom F6 professional field recorder with external RODE microphone. (c) Wildlife Acoustics Song Meter SM4 autonomous recorder powered by rechargeable NiMH batteries. All devices were mounted at consistent heights (1.5–2.0 m) to ensure comparable whole-room acoustic capture.

Zoom recorders were powered using portable rechargeable battery packs, while the Song Meter operated on four rechargeable D-size NiMH batteries, enabling uninterrupted long-duration recording. Microphones were mounted at uniform heights of 1.5 to 2.0 m above the litter surface to capture whole-room ambient sound at consistent gain settings. Recordings were acquired at a sampling rate of 44.1 kHz with 16-bit resolution and stored in WAV format to support high-fidelity signal processing and archival access.

### 2.3.3 Video recording and optical flow capture

Overhead GoPro Hero 13 cameras mounted in fixed positions within each room captured continuous video at 1080p resolution and 30 frames per second. Video collection commenced in week 5, as cameras could not be reliably operated during weeks 1 to 4 due to overheating in the high-temperature brooding environment required for early chick survival. From week 5 onward, AC-powered cameras maintained continuous recording through week 20. Figure 3 summarizes the video data acquisition setup, with an overhead GoPro Hero 13 camera and AC power supply enabling standardized, continuous flock monitoring across the 20-week study.

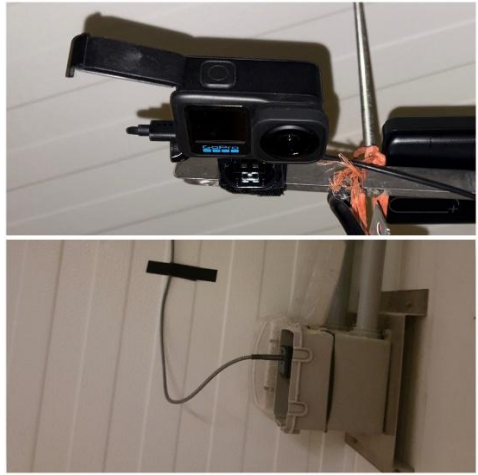

Figure 3. Video data acquisition hardware. (Upper) Overhead-mounted GoPro Hero 13 camera positioned to capture whole-room flock behaviour. (Lower) AC power supply enabling continuous camera operation across the 20-week study. Standardised mounting ensured temporal comparability across rooms.

Representative 15-minute video segments were subsequently selected for optical flow analysis, as described in Section 2.4.3. The absence of usable video data during the earliest brooding phase reflects current hardware constraints under extreme thermal conditions rather than a shortcoming of experimental design.

### 2.3.4 Environmental monitoring
Environmental conditions were monitored using two complementary systems providing both routine and contextual coverage. Wall-mounted DHT22 digital sensors, calibrated to ±2 percent accuracy, logged ambient temperature and relative humidity twice daily at 08:00 and 16:00 hours in each room. Supplementary air-quality parameters, including carbon dioxide concentration, volatile organic compounds, and particulate matter indices, were recorded once daily using portable multi-sensor arrays. These additional air-quality parameters were collected for contextual interpretation and future model extension, while temperature and relative humidity were the environmental variables used in the present analyses. Together, these measurements provided continuous environmental context essential for interpreting physiological and behavioural responses. Figure 4 shows the environmental monitoring apparatus, featuring a portable multi-sensor array that recorded temperature, relative humidity, $CO_2$, VOCs, and particulate matter.

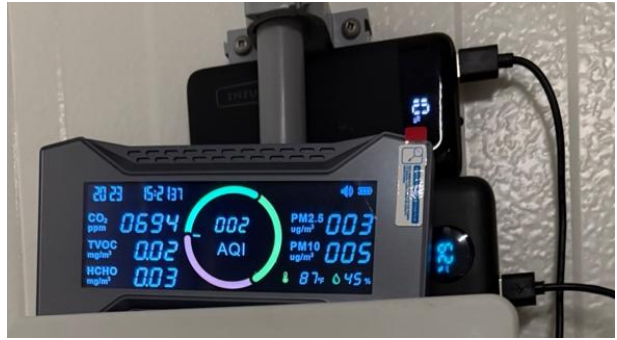

Figure 4. Environmental monitoring apparatus. Portable multi-sensor array used to record temperature, relative humidity, carbon dioxide concentration, volatile organic compounds, and particulate matter. Measurements were integrated with twice-daily wall-mounted DHT22 sensor logs to characterise room-level environmental conditions.

## 2.4 Data processing and quantitative analysis
### 2.4.1 Thermal feature extraction and analysis
Thermal images were processed using the FLIR Ignite platform to extract region-of-interest temperature metrics. For each image, minimum, maximum, and mean head and foot surface temperatures were recorded. Data from weeks 0 to 1, during which chick legs were not reliably visible, were excluded to prevent analytical bias. Weekly aggregates were calculated as mean and standard deviation across all valid images within each week for each room. Temporal patterns were visualised using line plots and boxplots stratified by developmental week. One-way analysis of variance was applied to test for weekly differences, with Tukey post-hoc comparisons identifying significant week-pair contrasts. Effect sizes were quantified using eta-squared.

### 2.4.2 Acoustic feature extraction and temporal validation
Raw audio recordings from Room 1 were processed using a standardised pipeline optimised for agricultural soundscapes. Preprocessing included spectral noise reduction using the *noisereduce* library, normalisation to consistent RMS intensity, and conversion to mono format. From each week, 10 to 12 one-minute clips were extracted during peak vocal activity between 06:00 and 09:00 hours, sampled from days 1, 4, and 7 to capture within-week developmental progression. Acoustic features were computed using the *librosa* library, including spectral centroid, spectral bandwidth, spectral rolloff, zero-crossing rate, RMS amplitude, and short-term energy. Weekly feature summaries were calculated as mean and standard deviation across all clips. Normality and homogeneity of variance were assessed using Shapiro-Wilk and Levene tests, respectively. When assumptions were violated, confirmatory non-parametric analyses using the Kruskal-Wallis test were conducted.

### 2.4.3 Optical flow analysis and behavioural quantification
Video data from Room 1 were processed using OpenCV version 4.5.1. Three 15-minute clips per week were selected from scheduled caretaker entry events on days 1, 4, and 7. Each clip was segmented into pre-entry baseline, caretaker entry, and post-entry recovery periods.

The Dense Inverse Search optical flow algorithm was applied to 30-second segments, generating pixel-wise motion vectors. Motion magnitude was aggregated across pixels to yield whole-flock movement intensity per segment. Weekly optical flow values were averaged by entry condition.

One-way analysis of variance assessed condition-dependent differences, and paired t-tests compared early (weeks 5 to 10) and late (weeks 11 to 20) developmental phases. Cohen's *d* was reported to quantify effect magnitude.

### 2.4.4 Cross-modal integration and correlation analysis
Weekly aggregated features from all modalities were compiled for 16 developmental weeks from weeks 5 to 20. Pearson correlation coefficients were computed for all pairwise feature combinations, corresponding to 10 features and yielding 45 unique comparisons. Multiple comparison correction was applied using the Benjamini-Hochberg false discovery rate procedure with a critical threshold of *q* < 0.05. Z-score normalisation was applied prior to visualisation to facilitate cross-modal comparison and identification of synchronous or divergent developmental patterns.

### 2.4.5 Statistical software and analytical environment
All analyses were performed in Python 3.9.7 using Pandas 1.3.4, NumPy 1.21.2, SciPy 1.7.1, and Statsmodels. Visualisations were generated with Matplotlib 3.4.3 and Seaborn 0.11.2. Statistical significance was defined as $p < 0.05$ for individual tests and $q < 0.05$ following multiple comparison correction. All models report test statistics, *p*-values, and effect sizes alongside descriptive summaries to support transparent interpretation of statistical and biological relevance.

## 2.5 Ethical approval and animal welfare
This study was conducted in strict accordance with the Canadian Council on Animal Care (CCAC) guidelines for humane animal use in research. All procedures were reviewed and approved by the Dalhousie University Animal Care and Use Committee (Dalhousie ACUC), Protocol Approval No. 2025-012. Birds were maintained under standard commercial rearing protocols without experimental manipulation beyond routine husbandry. The study was therefore classified as non-invasive observational research. Table 1 summarizes data collection across thermal imaging (all rooms, weeks 1–20), acoustic recording (Room 1, weeks 1–20), video optical flow (Room 1, weeks 5–20), and environmental monitoring (Room 1, weeks 1–20), with standardized quality control criteria applied to each modality.

## 2.6. Data Collection and Quality Control Summary

Table 1. Summary of multimodal data collection, temporal coverage, spatial coverage, sampling frequency, and quality control criteria across the 20-week study period.

| Modality | Temporal Coverage | Spatial Coverage | Sampling Frequency | Total Sample Count | Quality Control Criteria |
|---|---|---|---|---|---|
| Thermal Imaging | Weeks 1–20 | All rooms (1–5) | Every 3rd day | 120–140 images/room | Images free of blur, occlusion; standardized distance/angle; radiometric metadata validated |

| Modality | Temporal Coverage | Spatial Coverage | Sampling Frequency | Total Sample Count | Quality Control Criteria |
| --- | --- | --- | --- | --- | --- |
| Acoustic Recording | Weeks 1–20 | Room 1 (primary); Rooms 2–4 (archive) | 10–12 clips/week | 200–240 clips | Morning peak activity window (06:00–09:00); 1-min duration; spectral noise-reduction applied; mono normalized |
| Video (Optical Flow) | Weeks 5–20* | Room 1 | 3 clips/week | 48 total clips | Pre-entry (7 min), during entry (1–2 min), post-entry (7 min); days 1, 4, 7 of each week; downscaled 360p; DIS algorithm applied |
| Environmental Conditions | Weeks 1–20 | Room 1 (primary logger) | Twice daily (AM/PM) + daily supplements | 280+ readings | DHT22 sensors ±2% accuracy; calibrated wall-mounted units; supplementary air-quality indices; continuous coverage |

## 3. Results and Discussion

### 3.1 Environmental conditions as a contextual framework

Environmental measurements provided the backdrop against which developmental trajectories unfolded. Across the rearing period, ambient conditions showed predictable but nontrivial variation at both diurnal and seasonal scales. From June through October, temperatures exhibited consistent AM to PM differences, with afternoon readings typically higher than morning readings. This pattern reflects heat accumulation in enclosed housing and is representative of the operational realities of controlled poultry rooms, where even regulated systems contain diurnal structure driven by equipment cycles, stocking density, and thermal inertia. Other environmental variables such as $CO_2$ and ppm levels were treated as contextual covariates rather than primary outcome measures. Figure 5 illustrates seasonal temperature and humidity variation in Room 1, providing environmental context for multimodal developmental analysis.

Early summer conditions were characterised by higher afternoon temperatures, with peaks approaching 26–28°C. August was comparatively stable (mean 22.1°C, range 21.6–24.0°C), although it included the largest single excursion observed in the dataset, an AM decline of 4.2°C on 17 August. September and October remained moderate (mean 21.8 ± 2.1°C), yet still displayed intermittent daily oscillations exceeding 2°C. These fluctuations were not pathological, but they were sufficient to test the birds' developing capacity to buffer internal state against environmental variation.

Relative humidity varied inversely with temperature, consistent with closed housing environments in which warmer air typically corresponds to lower relative saturation. The temperature–humidity relationship was also supported by the correlation structure later observed in cross-modal analysis (temperature vs relative humidity, $r = -0.63$, $q = 0.043$). Importantly, environmental variation was not treated as an outcome in itself, but as contextual input that can modulate physiological signals and behavioural responses. A key value of this dataset is that it captures development under variable but realistic housing conditions rather than idealised constancy. This matters for precision livestock farming, where algorithms trained only under stable conditions often fail when confronted with normal environmental noise.

Two interpretation principles follow. First, environmental variation should not automatically be equated with poor welfare. Moderate variation is ubiquitous and may be well tolerated, especially as birds mature. Second, even when environmental conditions remain within acceptable ranges, they can still influence how biological signals present, particularly for acoustic features. This was evident in the selective humidity–acoustic correlations reported later. In this sense, environmental monitoring plays an enabling role: it prevents misattribution of environmentally induced shifts to intrinsic developmental change.

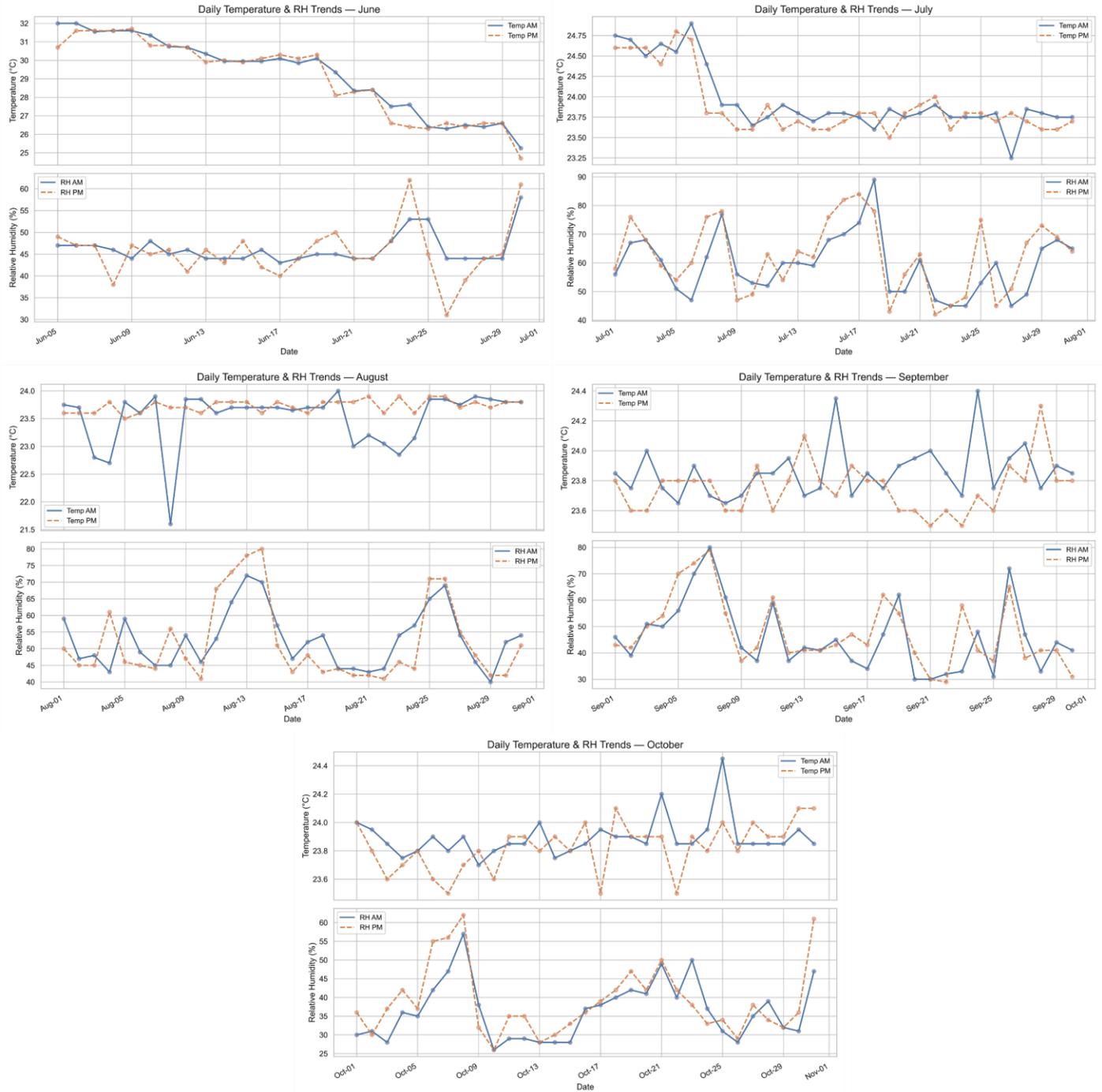

Figure 5. Ambient environmental conditions during rearing. Daily morning (08:00) and afternoon (16:00) temperature and relative humidity recorded in Room 1 from June to October. Seasonal variation is evident, with higher temperatures in early summer and more stable conditions in August. Environmental context informed interpretation of multimodal developmental trajectories.

**3.2 Thermoregulatory maturation: quantitative evidence and developmental timeline**
Thermal imaging captured systematic age-related change in peripheral surface temperatures, consistent with progressive maturation of thermoregulatory capacity. Head surface temperature exhibited significant weekly differences (F = 1.96, p = 0.028, η² = 0.38), indicating a moderate developmental effect. Tukey post-hoc analysis identified a significant contrast between week 2 and week 5 (p = 0.02), suggesting that a discernible adjustment phase occurred early, during the transition from intensive brooding dependency toward increased physiological autonomy. Table 2 reports ANOVA results and significant Tukey post-hoc comparisons for weekly head and foot surface temperatures, including F-statistics, p-values, effect sizes (η²), and week-pair differences. Beyond this early contrast, head temperatures showed increasing stability, with values plateauing by approximately week 10. Figure 6 shows head and foot temperature stabilization by weeks 7-10, indicating thermoregulatory maturation.

Foot surface temperature showed a stronger developmental pattern than head temperature. Weekly differences were highly significant (F = 3.22, p < 0.001, η² = 0.51), indicating that age accounted for a substantial portion of variance. Tukey comparisons detected significant differences between week 2 and week 5 (p < 0.001) and between week 2 and week 7 (p = 0.0102). After the early period, foot temperatures stabilised, with later weeks exhibiting narrower variation and a more consistent thermal profile. This is biologically plausible because peripheral surface temperature is strongly influenced by vasomotor control. The developmental increase and subsequent stabilisation is consistent with improving peripheral perfusion and maturing autonomic regulation. Two points deserve emphasis. First, the head and foot did not mature identically. Head temperatures exhibited a moderate developmental effect and stabilised later than the foot. Foot temperatures demonstrated a larger effect size and reached a stable regime earlier. This is a useful reminder that "thermoregulatory maturation" is not a single event, but a staged process distributed across body regions with different vascular architecture and heat exchange roles.

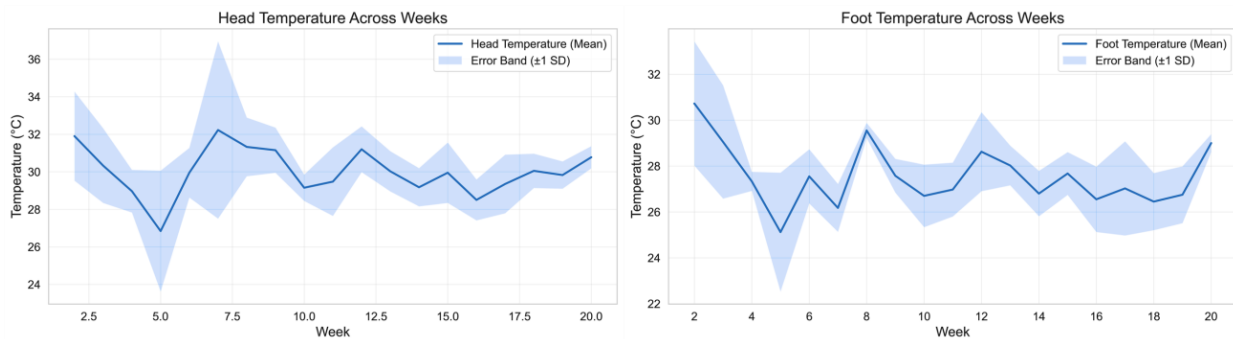

Figure 6. Thermoregulatory development across early life. Weekly mean head and foot surface temperatures from hatch to 20 weeks. Error bars represent standard deviation. Head temperatures stabilize by approximately week 10, while foot temperatures stabilize earlier (weeks 7–8), indicating maturation of central and peripheral thermoregulation.

Table 2. One-way ANOVA results and significant Tukey post-hoc pairwise comparisons for weekly head and foot surface temperatures across development. Reported values include F-statistics, p-values, effect sizes (η²), and week-pair contrasts showing statistically significant differences.

| Region | ANOVA F-value | ANOVA p-value | Effect Size (η²) | Significant Tukey Comparisons |
|---|---|---|---|---|
| **Head** | 1.96 | 0.028 | 0.38 | week 2 vs week 5 |

| Region | ANOVA F-value | ANOVA p-value | Effect Size (η²) | Significant Tukey Comparisons |
|---|---|---|---|---|
| **Foot** | 3.22 | <0.001 | 0.51 | week 2 vs week 5; week 2 vs week 7 |

Second, the observed stabilisation timeline appears slower than some classical descriptions of early thermoregulatory competence (29). Rather than interpreting this as delayed development, the more plausible explanation in this dataset is environmental variability. Daily and seasonal fluctuations provide repeated perturbations, and the measured surface temperatures track both maturation and context. The strong coupling of extremity temperature patterns with ambient curves, including cooler-period reductions consistent with vasoconstriction and warmer-period elevations, supports this interpretation. Under variable conditions, stabilisation is better understood as increasing robustness rather than a simple switch from immature to mature. In practical terms, robust thermoregulation in commercial settings is not the ability to maintain a constant surface temperature, but the capacity to regulate within tolerable bands while environmental inputs fluctuate.

Notably, the thermal profiles did not show patterns typically associated with severe or sustained dysregulation. While thermal imaging alone cannot diagnose welfare status, the absence of erratic multi-week instability provides convergent support for a generally coherent developmental course in this controlled setting. Future studies incorporating independent biomarkers would be required to translate these thermal trajectories into validated welfare inference.

### 3.3 Vocal maturation: acoustic signatures of developmental and social change
Acoustic feature trajectories showed clear age-related shifts consistent with vocal maturation and changes in flock-level communication. Frequency-related features exhibited systematic decline across development. Spectral centroid differed significantly across weeks ($F = 13.97$, $p < 0.001$), and spectral bandwidth also showed significant age effects ($F = 9.96$, $p < 0.001$). Spectral rolloff displayed a similarly strong developmental signal ($F = 17.99$, $p < 0.001$). Together, these shifts indicate a transition from higher-frequency, broader, and more variable vocalisations typical of young chicks toward lower-frequency and more stable spectral structure later in development (30, 31). Energy-related features moved in the opposite direction. RMS amplitude increased significantly with age ($F = 17.22$, $p < 0.001$), and short-term energy also differed across weeks ($F = 6.30$, $p < 0.001$). Table 3 reports one-way ANOVA results for age-related differences in acoustic spectral features across developmental weeks. This combination of decreasing frequency metrics and increasing energy metrics is consistent with a maturation process in which vocal production becomes stronger and more sustained while spectral content becomes more structured (12, 31). Mechanistically, such trajectories align with anatomical and neuromuscular development of the vocal apparatus, as well as changing behavioural drivers (32). Early calls are often dominated by contact and distress components, which can be high-frequency and less energetically sustained (33). Later communication increasingly reflects social organisation and routine flock-level signalling, which tends to be lower-frequency and more stable (34).

The dataset also demonstrates a useful methodological property: the acoustic pipeline was sensitive to short-lived disturbances that were confirmed by husbandry logs. Peaks in RMS amplitude and short-term energy occurred around weeks 11–13, coinciding with documented changes in housing conditions due to maintenance events. A pronounced anomaly in week 19 aligned with an equipment disturbance. These events serve two purposes. First, they confirm that the acoustic system captured genuine perturbations rather than generating smooth trajectories by artefact. Second, they highlight the importance of event logging for interpretability. Without metadata, anomalies risk being misclassified as biological stress signatures. With metadata, they become a feature, demonstrating system responsiveness.

Environmental modulation of acoustic features emerged as one of the strongest cross-modal signals in the dataset. Relative humidity was significantly positively correlated with zero-crossing rate ($r = 0.70$, $q = 0.014$) and spectral centroid ($r = 0.65$, $q = 0.031$) after false discovery rate control. The magnitude of these associations is moderate, not deterministic. This is important: the developmental trajectory is not merely a reflection of humidity, but humidity appears to modulate spectral properties in a measurable way. A plausible interpretation is that humidity influences respiratory mechanics and behaviour, which in turn affects vocal output. Another possibility is that humidity co-varies with other environmental dynamics that were not modelled, such as ventilation patterns or noise conditions. Because the study is descriptive, causal explanations remain speculative, but the association is robust enough to justify treating humidity as a meaningful contextual covariate in future AI-enabled acoustic monitoring systems. Figure 7 shows z-score-normalized acoustic feature trajectories from weeks 1–20, demonstrating significant age-related shifts ($p < 0.001$) consistent with vocal maturation, with anomalies corresponding to documented disturbances.

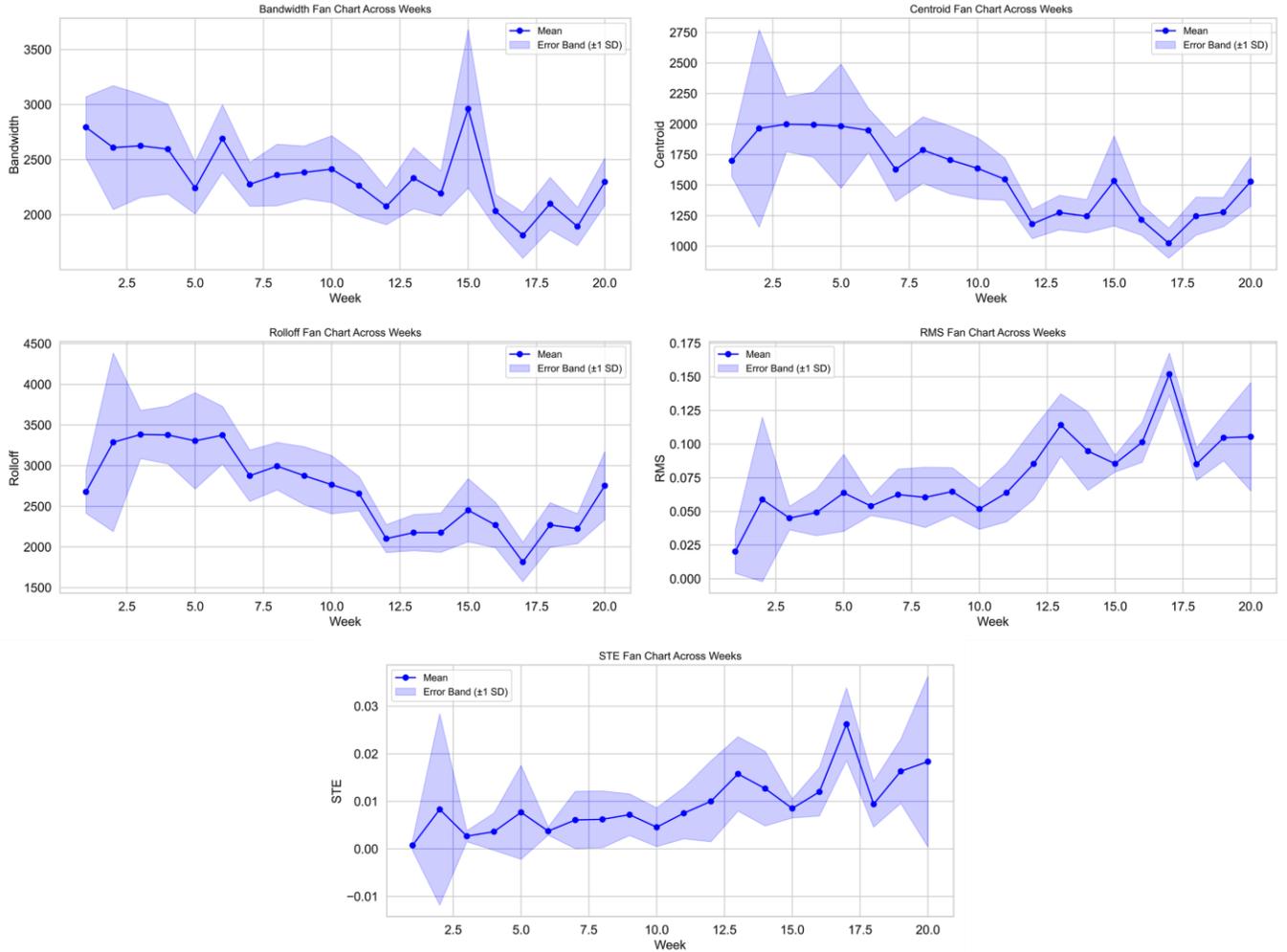

Figure 7. Acoustic feature trajectories across development. Z-score-normalized weekly trends in spectral bandwidth, spectral centroid, RMS amplitude, spectral rolloff, and short-term energy from weeks 1 to 20. All features show significant age-related shifts ($p < 0.001$), consistent with vocal maturation and changing social communication. Isolated anomalies correspond to documented non-routine disturbances.

Table 3. One-way ANOVA results for age-related differences in acoustic spectral features across developmental weeks.

| Spectral Feature | F-statistic | P-value |
| --- | --- | --- |
| Bandwidth | 9.96 | 2.22e-20 |
| Spectral Centroid | 13.97 | 1.07e-27 |
| RMS amplitude | 17.22 | 6.44e-33 |
| Short-term energy (STE) | 6.30 | 1.14e-12 |
| Spectral rolloff | 17.99 | 4.63e-34 |

Overall, acoustic features provide a coherent developmental signature that reflects intrinsic maturation and selective environmental modulation. Importantly, the weak behavioural correlations reported later indicate that vocal maturation is not simply a proxy for movement habituation. This strengthens the argument that acoustic monitoring adds complementary information rather than redundancy.

**3.4 Behavioural development: habituation as a quantifiable response axis**
Optical flow analysis provided an objective measure of flock movement around routine caretaker entry. Across the three entry conditions (before, during, after), optical flow differed strongly (F = 89.40, p < 0.001, $\eta^2$ = 0.86), indicating that caretaker condition accounted for a large proportion of movement variance. Post-hoc tests showed that movement during caretaker presence was significantly higher than both baseline and post-entry periods (during vs before: p = 0.0064; during vs after: p = 0.0267), while baseline and post-entry did not differ (p = 0.8497). This pattern suggests a stimulus-linked movement response that resolves quickly after the stimulus ends. In behavioural terms, it is consistent with a transient reaction to human presence rather than a prolonged state shift.

A second result speaks to developmental change. A paired comparison between early and late developmental periods showed a significant decline in movement reactivity over time (t = 28.12, p = 0.00126). This provides quantitative support for progressive habituation to routine caretaker entry. In early weeks, the flock showed stronger movement response to entry; later, the response amplitude narrowed, consistent with reduced novelty or reduced fear response under repeated predictable exposure (35). Table 4 summarizes statistical results for optical flow responses to caretaker entry, including ANOVA and post-hoc tests across developmental stages.

Two interpretive cautions are essential. First, optical flow is a flock-level motion metric, not a direct measure of affective state (36). Reduced movement response can reflect habituation, but it can also reflect changes in baseline activity, changes in flock density distribution, or even changes in camera geometry (37). The strength of the caretaker-condition effect and the stability of the pre/post baseline in the same clips supports a habituation interpretation, but formal welfare inference would require independent behavioural scoring or validated fear tests.

Second, optical flow is sensitive to non-routine disturbances. The dataset contains notable spikes during weeks 11–12 and week 19 linked to maintenance and equipment disruptions. These events are not nuisances. They demonstrate that optical flow is responsive to environmental perturbations that are behaviourally salient to the flock. From a precision monitoring perspective, this is an asset. A mature system could use context-aware anomaly detection to distinguish routine entry responses from atypical disturbances, providing early warning for management events or equipment failures. The key takeaway is that optical flow captures a behavioural response axis that changes across development and is strongly stimulus-linked. Its weak correlation with thermal and acoustic domains later suggests that it reflects a partially independent developmental process, likely involving learning and repeated exposure rather than purely physiological maturation. Figure 8 shows weekly optical flow magnitude before, during, and after caretaker entry (weeks 5–20), demonstrating pronounced early reactivity that declines in later weeks, consistent with behavioural habituation.

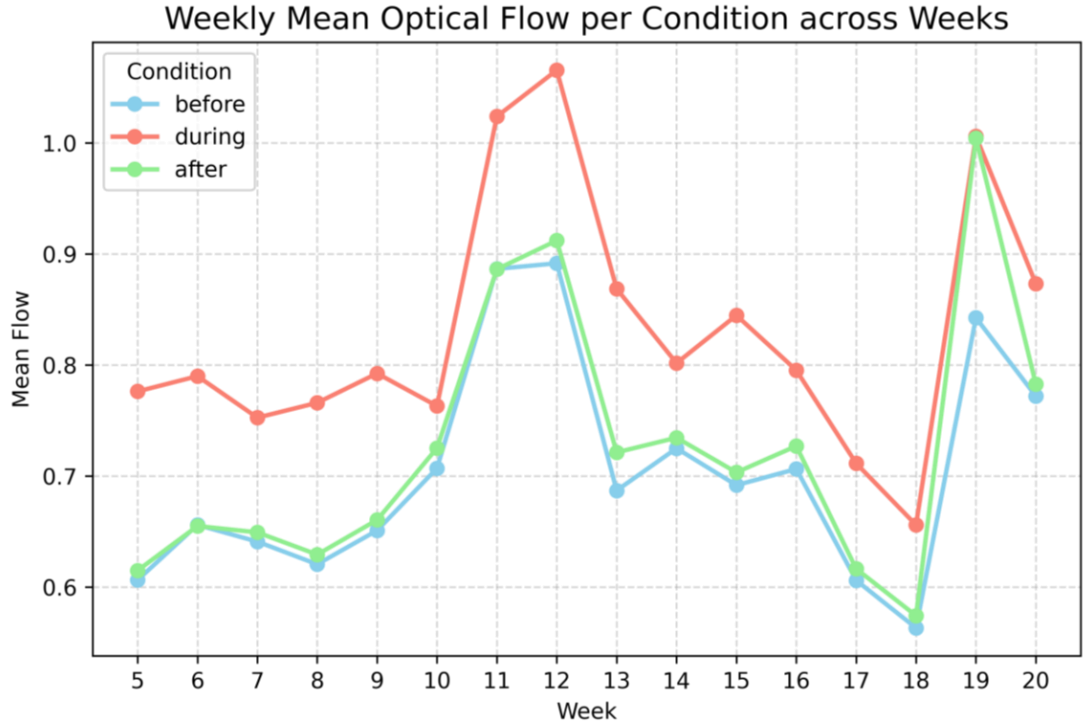

Figure 8. Behavioural response to caretaker entry. Weekly mean optical flow magnitude before, during, and after routine caretaker entry from weeks 5 to 20. Early weeks show pronounced reactivity to human presence, while later weeks exhibit reduced differential response, consistent with behavioural habituation.

Table 4. Statistical summary of optical-flow-based flock movement responses to routine caretaker entry across developmental stages.

| Analysis | Comparison | Test Statistic | p-value |
| --- | --- | --- | --- |
| One-way ANOVA | Before vs During vs After caretaker entry | f = 89.40 | 2.29e-13 |
| Tukey post-hoc | During vs Before | - | 0.0064 |
| Tukey post-hoc | During vs After | - | 0.0267 |
| Tukey post-hoc | Before vs After | - | 0.8497 |
| Paired t-test | Early weeks vs Late weeks | t = 28.12 | 0.00126 |

### 3.5 Integrated multimodal developmental trajectories

Visual integration of z-score-normalised trajectories provides an intuitive overview of how modalities evolve together while retaining distinct dynamics. In the integrated panel combining head temperature, spectral centroid, and baseline optical flow, all three show broad developmental movement toward stability or lower reactivity, but with different shapes. Head temperature stabilises relatively early. Spectral centroid declines more continuously. Baseline optical flow decreases in early development and then exhibits fluctuations around a lower set-point. The non-identical curves are informative. They imply that different biological systems mature on different schedules, and that multimodal designs capture complementary maturational signals rather than a single latent variable. Figure 9 shows z-score-normalized multimodal developmental trajectories across thermal, acoustic, behavioural, and environmental indicators, highlighting coordinated patterns with seasonal environmental variation.

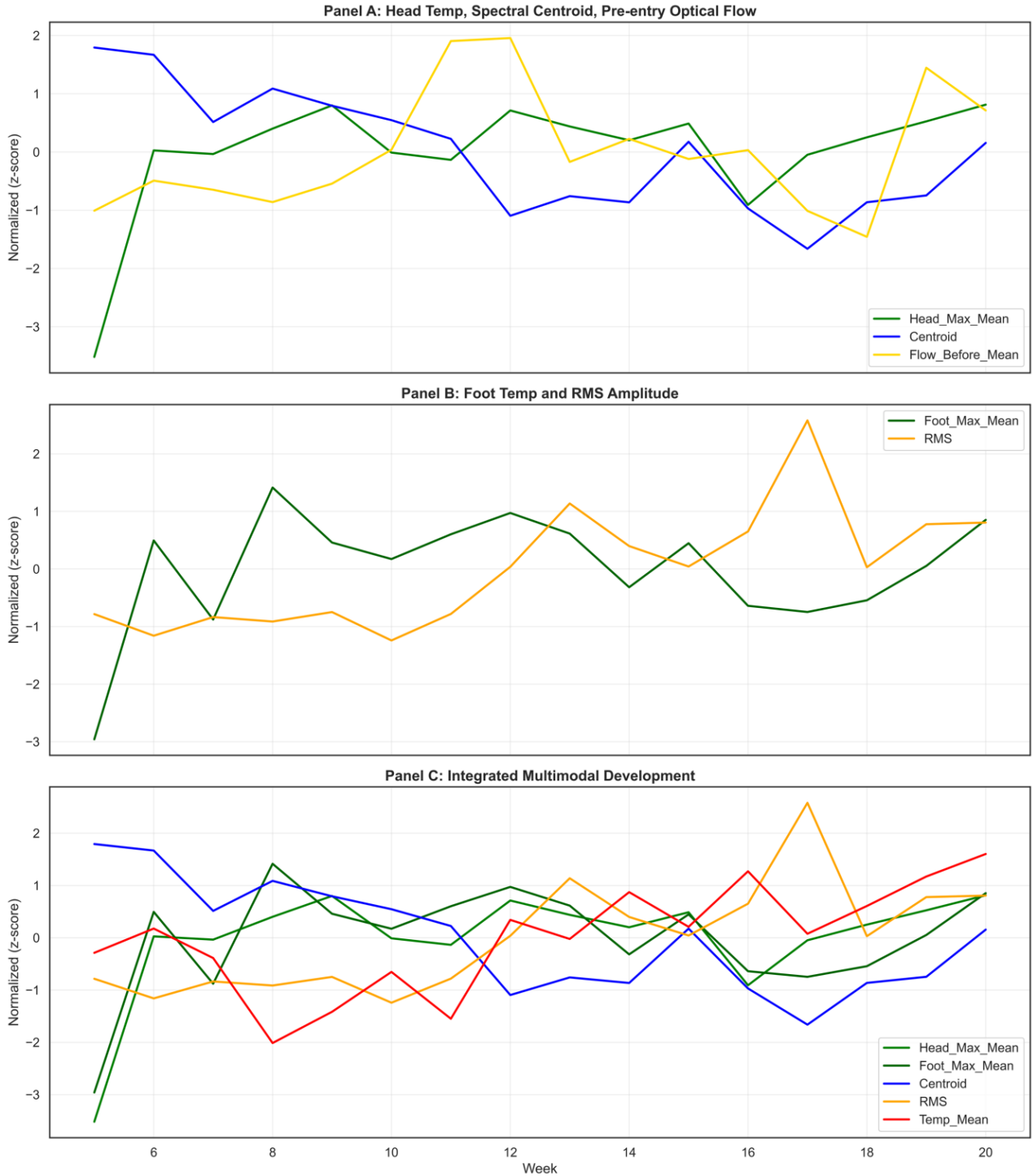

Figure 9. Integrated multimodal developmental trajectories. Z-score-normalized weekly features illustrating coordinated but modality-specific developmental patterns. (A) Head surface temperature, spectral centroid, and baseline optical flow. (B) Foot surface temperature and RMS acoustic amplitude. (C)

Combined thermal, acoustic, behavioral, and environmental trends highlighting developmental trajectories alongside seasonal environmental variation.

A second integrated panel combining foot temperature and RMS amplitude shows broadly parallel increases followed by stabilisation. This alignment is plausible, because both peripheral thermal stability and vocal energy may scale with growth, strength, and increasing social communication. However, parallelism should not be mistaken for equivalence. The later correlation analysis shows that thermal and acoustic domains do not uniformly co-vary in a way that supports strong cross-modal prediction. The visual overlap is useful as descriptive synthesis, but it does not replace quantitative association testing.

A third integrated panel that includes ambient temperature illustrates an important property of real-world sensing. Environmental signals oscillate with seasonal and operational patterns, while animal-based indicators often follow smoother developmental arcs. In this dataset, animal trajectories do not simply mirror environmental temperature. This suggests that developmental progression is buffered and that physiological maturity increases resilience to environmental noise. For PLF applications, this has practical consequences: models that ingest environmental data must be designed to avoid misattributing seasonal oscillation to biological change, and conversely, models that ignore environmental context risk false alarms when environmental conditions shift. Integrated visualisation is not evidence of causation. Its value lies in triangulation: it suggests where modalities move together, where they diverge, and where quantitative analysis should focus. In this pilot, it motivated the correlation analysis reported below and supports the claim that multimodal sensing yields structured developmental signals, not incoherent noise.

### 3.6 Cross-modal association structure and evidence for multidimensional development

Correlation analysis was performed across weekly-aggregated features from weeks 5–20, with Benjamini–Hochberg false discovery rate control ($q < 0.05$) applied across 45 comparisons. Within-modality correlations were consistently strong, supporting internal measurement coherence. Optical flow metrics before, during, and after entry were highly intercorrelated ($r = 0.92$–$0.96$, all $q < 0.001$). Thermal measures across regions were strongly associated (head vs foot temperature: $r = 0.85$, $q < 0.001$). Acoustic feature relationships followed expected patterns: spectral centroid and zero-crossing rate were strongly positively correlated ($r = 0.93$, $q < 0.001$), while RMS amplitude showed significant negative associations with centroid ($r = -0.80$, $q = 0.001$) and with ZCR ($r = -0.72$, $q = 0.010$). These patterns are not incidental. They reflect a coherent acoustic maturation signal in which frequency structure shifts downward while energy increases. Cross-modal associations were selective rather than pervasive. The strongest cross-modal links involved environmental context and acoustic structure, with humidity significantly associated with ZCR and spectral centroid ($r = 0.70$ and $r = 0.65$, respectively, $q < 0.05$). Temperature and relative humidity were also significantly negatively correlated ($r = -0.63$, $q = 0.043$), consistent with expected environmental physics in closed housing. Table 5 reports significant Pearson correlations between thermal, acoustic, behavioral, and environmental features after Benjamini–Hochberg FDR correction ($q < 0.05$).

Equally important are the non-significant associations. Thermal and optical flow metrics were weakly correlated and did not survive correction ($r = 0.24$–$0.43$, all $q > 0.27$). Acoustic and optical flow metrics were also weak and non-significant ($r = -0.31$ to $0.12$, all $q > 0.50$). A moderate acoustic–thermal association (ZCR vs head temperature: $r = -0.52$) did not survive FDR correction (q = 0.12). These results support a key biological conclusion: developmental change in thermoregulation, vocal structure, and behavioural habituation are partially independent. They can unfold concurrently over time without implying strong cross-prediction. Figure 10 shows the cross-modal correlation heatmap of thermal, acoustic, behavioral, and environmental features (weeks 5–20), with asterisks indicating FDR-corrected significance (q < 0.05), demonstrating strong within-modality consistency and selective cross-modal associations.

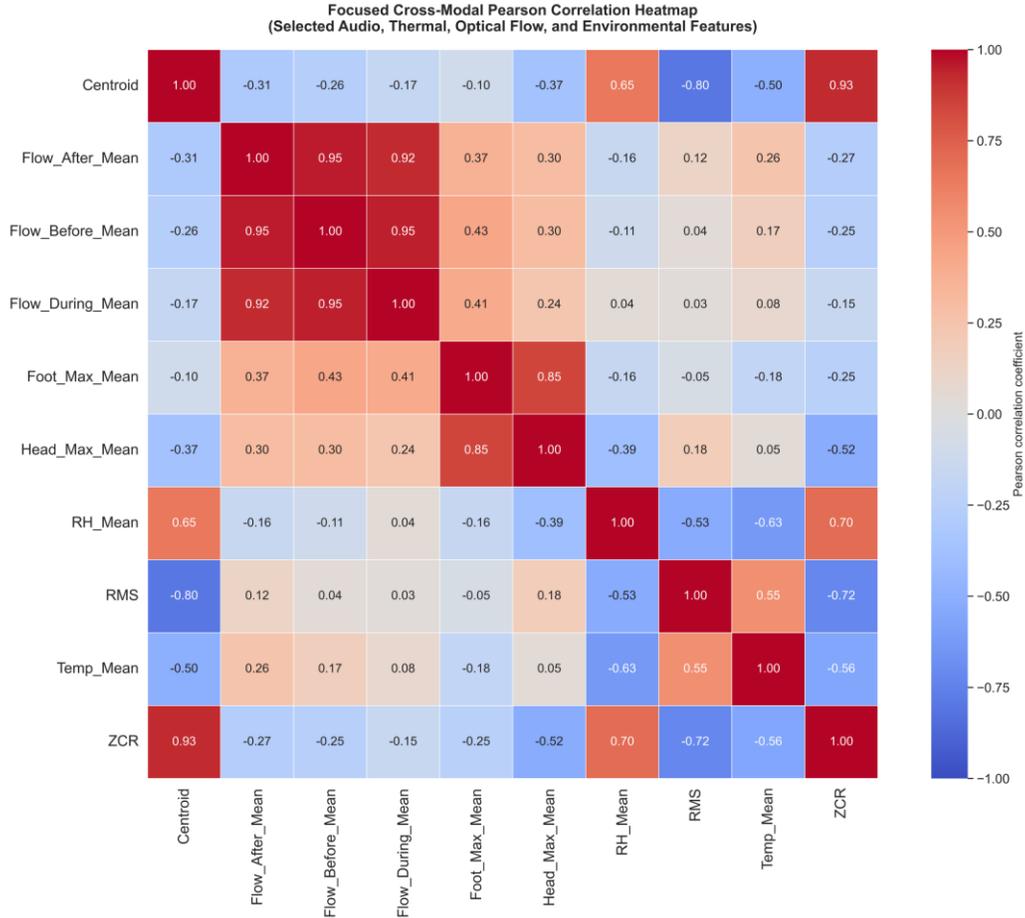

Figure 10. Cross-modal correlation structure. Heatmap of Pearson correlation coefficients between thermal, acoustic, behavioural (optical flow), and environmental features across weeks 5–20. Asterisks indicate significance after false discovery rate correction (q < 0.05). Strong within-modality correlations demonstrate internal consistency, while selective cross-modal associations highlight partially independent developmental domains.

This has direct implications for welfare-relevant monitoring. If modalities were redundant, single-modality systems would be sufficient. Instead, the dataset suggests that single-modality systems risk blind spots. Behavioural habituation to humans may progress independently of thermoregulatory maturity. Vocal maturation may progress independently of movement response. Environmental context may modulate vocal structure without strongly determining behavioural response. Multimodal sensing therefore offers a route to reduce these blind spots and to better localise which domain is changing when a deviation occurs.

From an engineering perspective, the selective association structure informs fusion strategy. It argues against naive early fusion that assumes all channels contain the same signal. Instead, it supports modular designs where each modality provides a domain-specific estimate and fusion occurs at the decision layer with explicit uncertainty handling.

Table 5: Significant Pearson correlations between thermal, acoustic, behavioural, and environmental features after Benjamini–Hochberg false discovery rate correction ($q < 0.05$).

| Correlation | Feature 1 | Feature 2 | r | p (uncorrected) | p_FDR | Significant ($q < 0.05$) |
|---|---|---|---|---|---|---|
| Within-Optical Flow | Flow Before | Flow After | 0.96 | < 0.001 | < 0.001 | Yes |
| Within-Optical Flow | Flow Before | Flow During | 0.95 | < 0.001 | < 0.001 | Yes |
| Within-Optical Flow | Flow During | Flow After | 0.92 | < 0.001 | < 0.001 | Yes |
| Within-Acoustic | Centroid | ZCR | 0.93 | < 0.001 | < 0.001 | Yes |
| Within-Acoustic | RMS | Centroid | −0.80 | < 0.001 | 0.001 | Yes |
| Within-Acoustic | RMS | ZCR | −0.72 | 0.001 | 0.010 | Yes |
| Within-Thermal | Head Temp | Foot Temp | 0.85 | < 0.001 | < 0.001 | Yes |
| Cross-Modal | ZCR | RH (Humidity) | 0.70 | 0.003 | 0.014 | Yes |
| Cross-Modal | Centroid | RH (Humidity) | 0.65 | 0.006 | 0.031 | Yes |
| Cross-Modal | Temp | RH (Humidity) | -0.63 | 0.010 | 0.043 | Yes |
| Cross-Modal | ZCR | Head Temp | -0.52 | 0.037 | 0.120 | No |
| Cross-Modal | Thermal | Optical Flow | 0.24-0.43 | > 0.09 | > 0.27 | No |
| Cross-Modal | Acoustic | Optical Flow | -0.31 to 0.12 | > 0.24 | > 0.50 | No |

### 3.7 Developmental staging and welfare-relevant synthesis

When integrated across modalities, the dataset supports a staged developmental narrative that is useful for interpretation, even though welfare validation is not possible in the absence of independent biomarkers. Figure 11 shows a conceptual synthesis of multimodal development across early, mid, and late phases.

**Early phase (weeks 1–6).** This period is characterised by higher variability in thermal measures, higher-frequency acoustic structure, and strong movement response to caretaker entry once video is available (from week 5). Conceptually, this phase reflects physiological immaturity and high sensitivity to environmental and social conditions. From a welfare monitoring perspective, the early phase is the period in which deviations may carry disproportionate consequences because buffering capacity is limited.

**Mid phase (weeks 7–13).** This period corresponds to stabilisation of peripheral thermal profiles, continued downward shift in acoustic frequency features, and gradual attenuation of behavioural response to routine entry. It also overlaps with documented environmental variability and husbandry events that produced identifiable anomalies in both acoustic and optical flow signals. The key insight is that the sensor system can discriminate between long-term developmental change and short-lived events when event logging is available.

**Late phase (weeks 14–20).** This period shows stable thermoregulatory patterns, stabilised acoustic profiles, and reduced movement reactivity to routine entry, consistent with increasing behavioural stability and physiological robustness approaching sexual maturity. This phase may represent the most tractable period for AI calibration in commercial settings because baseline variability is lower and developmental trajectories are closer to asymptote.

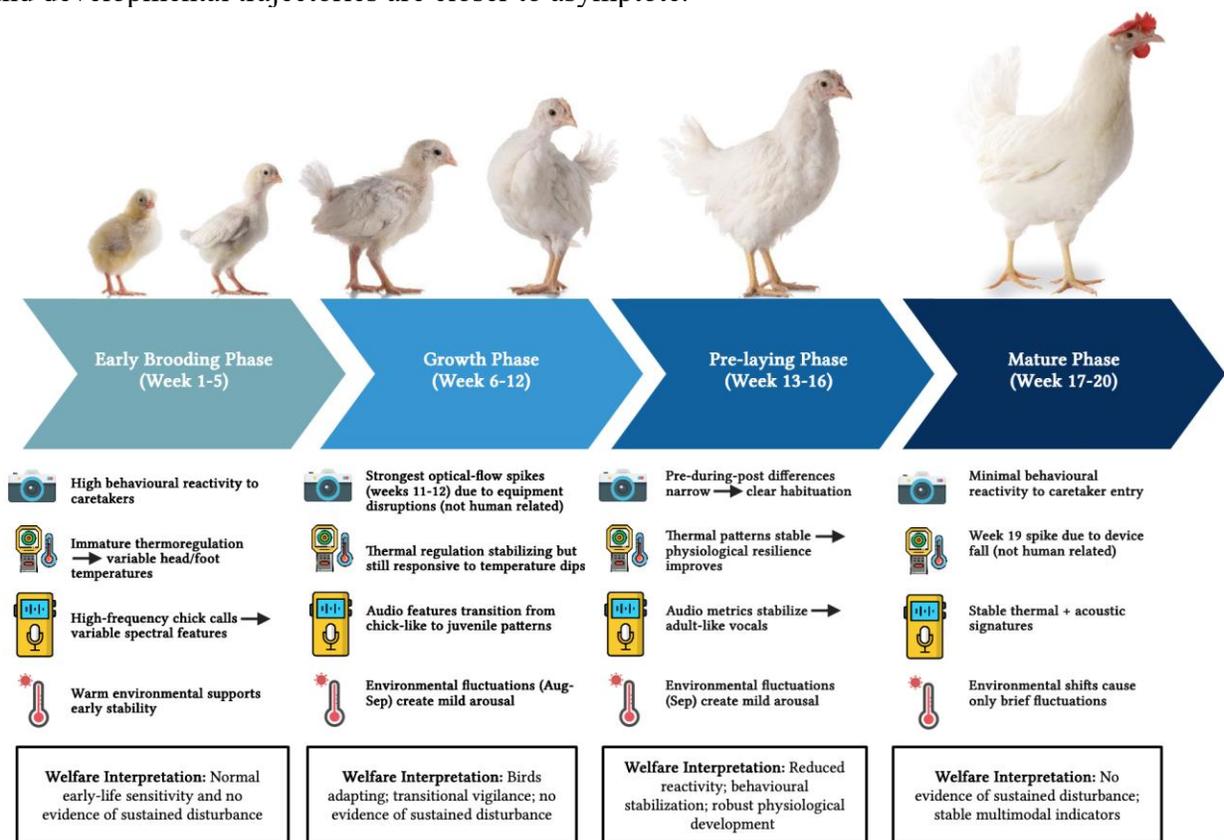

Figure 11. Conceptual synthesis of multimodal development in laying hens. Schematic representation of early (weeks 1–6), mid (weeks 7–13), and late (weeks 14–20) developmental phases, illustrating how thermal, acoustic, behavioural, and environmental indicators evolve in partially coordinated but distinct trajectories across early life.

This staging is not a welfare claim. It is a synthesis of developmental structure that can guide future study design. For example, if future work introduces validated fear tests or physiological biomarkers, these phases provide a rational framework for sampling design and for interpreting biomarker change relative to multimodal sensing trajectories.

### 3.8 Methodological scope and interpretive constraints

Several limitations define the appropriate scope of inference.

First, the study is descriptive and does not include independent welfare indicators such as corticosterone, heterophil-to-lymphocyte ratios, heart-rate variability, validated behavioural welfare scoring, or clinical outcomes. Therefore, sensor-derived patterns cannot be validated as welfare states, and interpretations must remain welfare-relevant rather than welfare-diagnostic.

Second, audio and video analyses were intentionally restricted to one representative room due to manual annotation demands, while thermal and environmental monitoring spanned all rooms. This design is appropriate for a feasibility pilot but limits generalisability and prevents formal cross-room inference for audio and behaviour.

Third, audio and video analyses were clip-based rather than continuous. Selected clips provide high-quality sampling across weeks, but rare events or transient welfare deviations may be missed. Fourth, video data were unavailable during weeks 1–4 due to camera overheating in brooding conditions, limiting behavioural analysis for the earliest phase.

Finally, analyses were based on weekly aggregation and room-level features. Individual-level variability and non-linear dynamics are not captured. These constraints do not undermine feasibility conclusions, but they strongly delimit claims about welfare, causation, and prediction.

### 3.9 Interpretation in light of literature and theory

The thermal trajectories are consistent with the ontogeny of avian thermoregulation, where peripheral vascular control and heat exchange evolve with growth and autonomic maturation (38). The stronger developmental effect in foot temperature compared with head temperature is biologically expected because extremities are highly sensitive to vasomotor regulation and environmental coupling.

Acoustic trajectories align with known developmental change in vocal production and social communication, where spectral structure shifts and energy increases as birds mature and social organisation stabilises (30, 31). The consistency across multiple acoustic descriptors and their internal correlation structure strengthens confidence that the system captured biological change rather than measurement noise.

Behavioural trajectories captured by optical flow are consistent with repeated exposure effects and habituation to routine caretaker entry (39, 40). The strong condition effect and the significant early-to-late reduction support the interpretation of decreasing reactivity under repeated predictable exposure, but welfare inference should remain cautious without direct behavioural scoring or fear tests.

The multimodal correlation structure provides a more nuanced contribution. It demonstrates that while sensor streams are individually coherent, their relationships are selective. This supports a multidimensional view of development and argues for multimodal monitoring as an approach to reduce blind spots rather than to increase redundancy. The strongest cross-modal links involve environmental context modulating vocal spectral structure, a point that is especially relevant for AI systems, because acoustic classifiers can be confounded by environmental variation unless context is modelled.

## 3.10 Implications for precision livestock farming and future directions

This pilot establishes three outcomes that are relevant to precision livestock farming.

First, long-duration multimodal data collection in a controlled poultry facility is technically feasible. Thermal imaging at regular intervals, sustained acoustic recording, continuous video recording from week 5 onward, and routine environmental logging were implemented over 20 weeks with interpretable outputs. The practical issues encountered, such as camera overheating during brooding, are themselves valuable design information for deployment.

Second, the sensor streams yielded biologically interpretable, internally coherent developmental trajectories. Strong within-modality consistency indicates that each modality captured structured developmental information rather than uncorrelated noise.

Third, cross-modal analysis is feasible even under pilot constraints, enabling quantification of which modalities are linked and which represent independent axes. This is precisely the information needed to design scalable, deployment-ready systems.

Future work should focus on five priorities:
1. synchronising multimodal sensing with independent welfare biomarkers and validated behavioural scoring to establish welfare relevance and predictive validity;
2. automating audio and video feature extraction to remove annotation bottlenecks and support multi-room scaling;
3. extending to multi-site and multi-strain datasets to evaluate generalisability;
4. introducing controlled perturbations to test specificity and sensitivity of multimodal responses to defined welfare challenges; and
5. developing machine-learning models that explicitly incorporate context, including environmental conditions and event logs, to reduce false alarms and improve robustness.

## 4. Conclusions

Early-life development in laying hens is a complex and inherently multidimensional process involving concurrent maturation of thermoregulatory, vocal, behavioural, and social systems. These domains evolve in parallel but are not redundant, and their interactions shape the biological and welfare trajectories that persist into adulthood. This pilot study demonstrates that such diverse developmental dimensions can be captured simultaneously and coherently through multimodal sensor integration, establishing both technical feasibility and biological plausibility for a new generation of welfare-relevant monitoring approaches in poultry systems.

Across a 20-week rearing period, thermal imaging revealed progressive stabilization of peripheral thermoregulation, with foot surface temperature showing a strong developmental effect ($\eta^2 = 0.51$). Acoustic analysis documented systematic age-related shifts from high-frequency, low-energy early-life calls toward lower-frequency, higher-energy vocalizations characteristic of mature social communication (all major spectral features $p < 0.001$). Optical-flow-based behavioural analysis quantified pronounced developmental attenuation of flock reactivity to routine caretaker entry, consistent with habituation under repeated non-aversive exposure. Importantly, these trajectories unfolded in coordinated yet non-identical temporal patterns. Quantitative cross-modal analysis revealed strong within-modality coherence alongside selective cross-modal associations, most notably between environmental humidity and acoustic features (r = 0.65–0.70), while thermal, acoustic, and behavioural domains remained largely independent after correction for multiple comparisons. This structure provides empirical support for the view that no single modality can adequately represent welfare-relevant development.

From a methodological perspective, this study accomplishes the core objectives of rigorous pilot research. It demonstrates that long-duration multimodal data collection is feasible in a commercial-scale research facility, documents internally consistent baseline developmental trajectories, and applies statistically conservative integration methods to quantify cross-modal relationships. Transient responses to non-routine environmental disturbances, detected independently in acoustic and behavioural signals and resolved thereafter, further validate that the sensing framework captures biologically meaningful variation rather than artefactual noise.

At the same time, the scope and limitations of this work must be clearly acknowledged. The study is descriptive rather than experimental, and no independent physiological or behavioural welfare biomarkers were collected. Detailed audio and video analyses were restricted to one representative room due to manual processing demands, and analyses were conducted at the room level using weekly aggregation. These constraints preclude causal inference, welfare validation, individual-level interpretation, or real-time deployment claims. Rather than weaknesses, these boundaries precisely define the study's contribution: proof of technical integratability and biological coherence, not proof of welfare prediction.

The pathway forward is therefore well defined. Future research should integrate independent welfare biomarkers synchronized with multimodal sensing, automate feature extraction to eliminate annotation bottlenecks, extend validation across multiple sites and strains, and incorporate controlled perturbations to test sensitivity and specificity. Supervised learning approaches trained on validated outcomes will be required to translate multimodal features into actionable decision support.

In broader context, this work reinforces a growing consensus in animal welfare science that welfare is multidimensional and cannot be reduced to a single physiological or behavioural axis. Multimodal monitoring does not replace clinical assessment or welfare scoring, but complements them by providing continuous, objective, high-frequency insight into developmental dynamics. This pilot establishes that such integration is technically achievable, biologically interpretable, and scientifically justified, providing a robust foundation for advancing precision livestock farming toward more responsive, evidence-based welfare management in laying hen production.

**Data availability statement**
The datasets generated and analyzed during this study are available from the corresponding author upon reasonable request due to large file sizes.

**Funding**

The authors sincerely thank the Natural Sciences and Engineering Research Council of Canada and Egg Farmers of Canada for their generous funding support for this project.